\documentclass[12pt]{article}
\usepackage{amsmath,amsfonts,amssymb}
\begin{document}
\thispagestyle{empty}
\newcommand{\bea}{\begin{eqnarray}}
\newcommand{\eea}{\end{eqnarray}}
\newcommand{\be}{\begin{eqnarray}}
\newcommand{\ee}{\end{eqnarray}}
\newcommand{\ben}{\begin{eqnarray*}}
\newcommand{\een}{\end{eqnarray*}}
\newcommand{\beq}{\begin{equation}}
\newcommand{\eeq}{\end{equation}}

\def\bJ{{\mathbb J}}
\def\bR{{\mathbb R}}
\def\bL{{\mathbb L}}
\def\bQ{{\mathbb Q}}
\def\bC{{\mathbb C}}
\def\bH{{\mathbb H}}
\newcommand{\non}{\nonumber}
\newcommand{\id}{\mathbb{I}}
\newcommand{\tr}{\mathop{\rm tr}\nolimits}
\newcommand{\sgn}{\mathop{\rm sign}\nolimits}
\newcommand{\diag}{\mathop{\rm diag}\nolimits}
\newcommand{\cf}{\ensuremath{\mathfrak{c}}}
\def\nn{\nonumber}
\def\tr{{\rm tr}\,}
\def\p{\partial}
\newcommand{\bde}{{\bf e}}
\renewcommand{\thefootnote}{\fnsymbol{footnote}}
\def\theequation{\arabic{section}.\arabic{equation}}
\def\a{\alpha}
\def\b{\beta}
\def\g{\gamma}
\def\d{\delta}
\def\dd{\rm d}
\def\e{\epsilon}
\def\ve{\varepsilon}
\def\z{\zeta}
\def\B{\mbox{\bf B}}\def\cp{\mathbb {CP}^3}

\newcommand{\h}{\hspace{0.5cm}}

\begin{titlepage}
\renewcommand{\thefootnote}{\fnsymbol{footnote}}
\begin{center}
{\Large \bf Finite-size effects of $\beta$-deformed ${\rm AdS}_5/{\rm CFT}_4$}
\vskip .5cm
{\Large \bf at strong coupling}
\end{center}
\vskip 1.2cm \centerline{\bf Changrim  Ahn$^1$, Diego Bombardelli$^2$, and Minkyoo Kim$^3$}

\vskip 10mm

\centerline{\sl $^1$Department of Physics and the Institute for the Early Universe} \centerline{\sl Ewha Womans University}
\centerline{\sl DaeHyun 11-1, Seoul 120-750, S. Korea}
\centerline{\tt ahn@ewha.ac.kr}
\vskip 1cm
\centerline{\sl $^2$Centro de F\'isica do Porto and Departamento de F\'isica e Astronomia}
\centerline{\sl Faculdade de Ci\^encias da Universidade do Porto}
\centerline{\sl Rua do Campo Alegre 687, 4169-007 Porto, Portugal}
\centerline{\tt diego.bombardelli@fc.up.pt}
\vskip 1cm
\centerline{\sl $^3$Department of Physics and Center for Quantum Spacetime}
\centerline{\sl Sogang University} \centerline{\sl Seoul 121-742, S. Korea}
\centerline{\tt mkim80@sogang.ac.kr}
\vskip 1cm

\baselineskip 18pt

\begin{center}
{\bf Abstract}
\end{center}
We compute both classical and quantum finite-size corrections at leading order in the strong coupling limit for the (dyonic) giant magnon in the Lunin-Maldacena background.
Based on the exact $S$-martix conjectured for the deformed theory, we generalize the L\"uscher formula to include twisted boundary conditions and show that the results match with those derived both by finite-size classical solutions of the giant magnon and by algebraic curve analysis.

\end{titlepage}

\newpage
\baselineskip 18pt

\vskip 0cm

\renewcommand{\thefootnote}{\arabic{footnote}}
\setcounter{footnote}{0}

\setcounter{equation}{0}
\section{Introduction}

Integrability discovered in the AdS/CFT duality between type IIB string
theory on $AdS_5\times S^5$ and ${\cal N}=4$ super Yang-Mills (SYM) theory
\cite{MGKPW} led to many exciting developments and to
understanding non-perturbative structures of both string and gauge theories
\cite{Review}.
This duality has been generalized to a one-parameter marginal deformation of SYM, the so-called $\beta$-deformed SYM theory,
which still preserves ${\cal N}=1$ supersymmetry \cite{LeiStr, MPSZ},
and even to a three-parameter deformed theory which has no supersymmetry
\cite{Frolov,FrRoTs}.
The deformed SYM theory is obtained by replacing the original
${\cal N}=4$ superpotential for the chiral superfields by:
\beq
W= ih\ {\rm tr}
(e^{i\pi\beta}\phi\psi Z-e^{-i\pi\beta}\phi Z\psi).
\eeq
The deformation breaks the supersymmetry down to ${\cal N}=1$ but still maintains
the conformal invariance in the planar limit to all perturbative orders
\cite{LeiStr, MPSZ,Zoubos}, since the deformation becomes exactly marginal
for real $\beta$ if
\beq
h{\overline h}=g_{\rm YM}^2,
\eeq
where $g_{\rm YM}$ is the Yang-Mills coupling constant.
When $\beta$ is real, this deformed SYM theory is dual to a type-IIB string theory
on the Lunin-Maldacena background \cite{LunMal}, which is obtained by a so-called TsT transformation.

In the weak coupling limit $\lambda\equiv g_{\rm YM}^2 N_{c}\ll 1$, various perturbative analysis of the deformed SYM has been studied \cite{FrRoTs} and, in
particular, anomalous dimensions for
the one and two magnon states in the $su(2)$ sector have been computed
up to four loops \cite{FSSZ}.
There have been several indications that the anomalous dimensions of the $\beta$-deformed SYM are exactly solvable.
Perturbative dilatation operators are mapped to some integrable spin chains \cite{
BerChe} and all-loop Bethe ansatz equations have been proposed \cite{BeiRoi}.
A first non-trivial check about the perturbative four-loop
anomalous dimension of the Konishi operator in the deformed gauge theory has been done recently in \cite{ABBN1} by computing it
from the L\"uscher formula \cite{Luscher,AJK,JL,BajJan} based on some twisted $S$-matrix elements.

Finite-size corrections for this and other operators of the deformed theory have been then obtained by using few different methods.
One way is to introduce ``operatorial'' twisted boundary conditions (BCs)
\cite{Arutyunov:2010gu},  another is to consider the untwisted Y-system
with twisted asymptotic conditions \cite{GroLev}.
Instead, our approach in this paper will be to combine both a Drinfeld-Reshetikhin
twisted $S$-matrix with ordinary twisted BCs \cite{ABBN2}.
In the developments of AdS/CFT duality, the $S$-matrix has been playing
an essential role \cite{Staudacher,S-matrix}.
This approach has been recently applied to compute next-to-leading order
L\"uscher (double wrapping) corrections to the vacuum of the three parameters non-supersymmetric deformed $AdS_{5}/CFT_{4}$ \cite{Frolov:2005iq,ABBN3} (see also \cite{deLeeuw:2012hp} for a recent generalization to orbifolds and deformations of the $AdS_5$ sector).

In the strong coupling regime, the string theory on this deformed background maintains
the classical integrability \cite{Frolov,BykFro}, and has identical excitations
such as giant magnons \cite{HM}, whose finite-size effects have been obtained by transforming the $AdS_5\times S^5$ background under a TsT transformation \cite{BykFro}:
\be
E-J&=&2g\sin\frac{p}{2}-\frac{8}{e^2}g\sin^3\frac{p}{2}\cos\Phi e^{-\frac{J}{g\sin p/2}}+\ldots\ ,
\label{BykFro}
\ee
where $g=\frac{\sqrt{\lambda}}{2\pi}$ and the effect of the deformation $\beta$ appears only through the phase $\Phi$:
\be
\Phi&=&\frac{2\pi(n_2-\beta J)}{2^{3/2}\cos^3\frac{p}{4}}.
\label{phBykFro}
\ee
Here $n_2$ corresponds to the untwisted boundary conditions of the isometric angles $\phi_{2}$ and is the integer closest to $\beta J$, such that $2\pi (n_{2}-\beta J)$ is restricted between $-\pi$ and $\pi$.
We recall that in the string classical limit one has $J\sim g\gg 1$
and the deformation parameter scales like $\beta\sim
1/g$. For the dyonic case, the second angular momentum $Q$ scales like $Q\sim g$.

Recently, a reanalysis of this calculation has led to a different result for the phase $\Phi$ \cite{AhnBoz,AhnBoz2} \footnote{This result was originally derived for the spectrum of the $CP^3_{\beta}$ giant magnon \cite{AhnBoz} and for the three-point correlation function of the $S^{5}_{\beta}$ giant magnon \cite{AhnBoz2} but it still holds for its energy since basically the same computation is involved.}.
For the case of the dyonic giant magnon, the finite-size effect turns out to 
be
\bea
E-J&=&\epsilon_Q(p)-\frac{16g^2\sin^4(p/2)}{\epsilon_Q(p)}\cos\Phi
\exp\left[-\frac{2\sin^2\frac{p}{2}\epsilon_Q(p)[J+\epsilon_Q(p)]}{Q^2+4g^2\sin^4\frac{p}{2}}\right],
\label{AhnBoz}\\
\Phi&=&2\pi (n_2-\beta J)+\frac{Q[J+\epsilon_Q(p)]\sin p}
{Q^2+4g^2\sin^4\frac{p}{2}},
\label{AhnBozi}
\eea
where $\epsilon_Q(p)$ is the dyonic dispersion relation 
\beq
\epsilon_{Q}(p)=\sqrt{Q^{2}+4g^2\sin^2\frac{p}{2}},
\label{disper1}
\eeq 
and $n_{2}$ now is allowed to be any integer number.
In the non-dyonic limit ($Q/\sqrt{\lambda}\to 0$), the phase $\Phi$ becomes 
\beq
\Phi=2\pi (n_2-\beta J)
\label{AhnBozii}
\eeq
which differs from (\ref{phBykFro}).
One of the main purposes of this letter is to confirm Eqs. (\ref{AhnBozi}) and (\ref{AhnBozii}) by calculating
L\"uscher $\mu$-term formula based on the twisted $S$-matrix and the twisted BCs.
This computes a shift in the energy due to
the finite-size of spatial length from the $S$-matrix
for all values of the `t Hooft coupling constant.
This method has been successfully applied to the undeformed AdS/CFT duality in \cite{AJK,JL,GSV,HatSuzi,BajJan,HatSuzii}.
Differently from the undeformed case, we will modify the formula to include the
twisted BCs.
We will also study a leading one-loop correction in the strong coupling regime
using the L\"uscher $F$-term formula and compare with the algebraic curve analysis.
\setcounter{equation}{0}
\section{Finite-size effects from the L\"uscher formulas}

It has been noticed that the three-parameter deformed Yang-Mills theory
can be described by a
Drinfeld-Reshetikhin twisted $S$-matrix with ordinary twisted BCs \cite{ABBN2}.
The twisted $S$-matrix is given by
\be
\tilde {\cal S}(p_{1},p_{2}) = F\, {\cal S}(p_{1},p_{2})\, F \,, \quad
{\cal S}(p_{1},p_{2})=S(p_{1},p_{2})\otimes S(p_{1},p_{2})
\label{drinfeldtwist}
\ee
where $S(p_{1},p_{2})$ is the $su(2|2)$ $S$-matrix \cite{S-matrix} and
the twist matrix $F$ is given by
\be
F = e^{i \gamma_{1} (h \otimes \id \otimes \id \otimes h -
\id \otimes h \otimes h \otimes \id )} \,,
\label{twistagain}
\ee
with a diagonal matrix $h$ given by
\be
h=\diag(\frac{1}{2},-\frac{1}{2},0,0) \,.
\label{hh}
\ee
The twisted BCs are imposed by a matrix $M$ which appears
in the definition of the (inhomogeneous) transfer matrix
\be
t(\lambda) = {\rm STr}_{a \dot a} M_{a \dot a} \tilde {\cal S}_{(a \dot a)
(a_1 {\dot a_1})}(\lambda,p_{1})
\ldots  \tilde {\cal S}_{(a \dot a)(a_N {\dot a_N})}(\lambda,p_{N}) \,,
\label{twistedBCagain}
\ee
where the matrix $M_{a \dot a}$ is given by
\be
M = e^{i (\gamma_{3}-\gamma_{2}) J h}
\otimes e^{i (\gamma_{3}+\gamma_{2}) J h}  \,,
\label{MAdSCFT}
\ee
and $J$ is the angular momentum charge which is related to the length of
spin chain by $J=L-N$.
We will restrict our analysis to the $\beta$-deformed case given by $\gamma_1=\gamma_2=\gamma_3\equiv2\pi\beta$.

\subsection{L\"uscher $F$-term and $\mu$-term formulas}

We propose that the L\"uscher $F$-term formula for a generic physical bound state with twisted BCs, is given by \footnote{The indexes $a, \dot a$ denote the $SU(2|2)\otimes SU(2|2)$ labels.}
\beq
\delta E^{F}_{(a\dot a)_{Q}}=-\int\frac{dq}{2\pi}\left(1-\frac{\epsilon_{Q}'(p)}{\epsilon_{1}'(q_{\star})}
\right)e^{-iq_{\star}J}\sum_{b,{\dot b},b',{\dot b'}}(-1)^{F_{b}+F_{\dot b}}
\left[M_{b'{\dot b'}}^{b{\dot b}}\left(\tilde{\cal S}_{(b{\dot b})(a\dot a)_{Q}}
^{(b'{\dot b'})(a\dot a)_{Q}}(q_{\star}(q),p)-1\right)\right].
\label{fterm}
\eeq

In the derivation of the $F$-term formula \cite{Luscher,JL,Janik}, 
there is a step where 
the integration contour is shifted from complex to real axis. 
When the $S$-matrix has a pole corresponding to a physical boundstate, 
the shift of contour can generate an extra term, which is
the so-called $\mu$-term:
\beq
\delta E^{\mu}_{(a\dot a)_{Q}}=-i\left(1-\frac{\epsilon_{Q}'(p)}{\epsilon_{1}'({\tilde q}_{\star})}
\right)e^{-i{\tilde q}_{\star}J}\sum_{b,{\dot b},b',{\dot b'}}(-1)^{F_{b}+F_{\dot b}}
\mathop{{\rm Res}}_{q={\tilde q}}
\left[ M_{b'{\dot b'}}^{b{\dot b}}\tilde{\cal S}_{(b{\dot b})(a\dot a)_{Q}}
^{(b'{\dot b'})(a\dot a)_{Q}}(q_{\star}(q),p)\right],
\label{muterm}
\eeq
where ${\tilde q}$ is the location of $S$-matrix the pole(s) and we use a short notation ${\tilde q}_{\star}=q_{\star}({\tilde q})$.
In the strong coupling limit, the $\mu$-term gives the leading classical contribution, while the $F$-term correspond to the first quantum finite-size correction. 

The L\"uscher corrections need only the $S$-matrix elements
which have the same incoming and outgoing
$SU(2|2)$ quantum numbers after scattering with a virtual particle.
In particular, we consider a bound-state of $Q$ $su(2)$ magnons in the physical particle state,
namely $(1\dot1)_{Q}$. It has momentum
$p$ and energy given by (\ref{disper1}), while the momentum of the virtual particle, $q_{\star}$, satisfies the following on-shell relation
\beq
q^{2}=-\epsilon_{1}^{2}(q_{\star}).
\label{disper}
\eeq
In this case, the twisted $S$-matrix elements can be written as
\beq
\tilde{\cal S}_{(b{\dot b})(1{\dot 1})_{Q}}
^{(b'{\dot b'})(1{\dot 1})_{Q}}=
\left[e^{i\pi\beta Q(h_{b}+h_{b'})}S_{b1_{Q}}^{b'1_{Q}}\right]\times
\left[e^{-i\pi\beta Q(h_{\dot b}+h_{\dot b'})}S_{{\dot b}\dot1_{Q}}^{{\dot b'}\dot1_{Q}}\right].
\eeq
Now, since the twisted BC matrix is a diagonal matrix which, in the case of $\beta$-deformation, becomes
\be
M = \id \otimes e^{4i\pi\beta J h}  \,,
\ee
then the sum in Eq.(\ref{fterm}) results to be
\beq
\sum_{b=1}^4
\left[(-1)^{F_b}e^{2i\pi\beta Q h_{b}}S_{b1_{Q}}^{b1_{Q}}\right]\times
\sum_{{\dot b}=1}^4\left[(-1)^{F_{\dot b}}
e^{2i(2J-Q)\pi\beta h_{\dot b}}S_{{\dot b}\dot1_{Q}}^{{\dot b}\dot1_{Q}}\right].
\label{summand}
\eeq

The explicit matrix elements are given by
\be
{S}^{\dot b\dot1_{Q}}_{\dot b\dot1_{Q}}(y^{\pm},X^{\pm})={S}^{b1_{Q}}_{b1_{Q}}(y^{\pm},X^{\pm})=S_0(y^{\pm},X^{\pm}){s}_b(y^{\pm},X^{\pm}),
\ee
where \cite{HatSuzi}
\beq
S_0^{2}(y^{\pm},X^{\pm})=\sigma_{\rm BES}(y^{\pm},X^{\pm})^2\frac{X^{+}}{X^{-}}\left(\frac{y^{-}}{y^{+}}\right)^{Q}\frac{y^+-X^-}{y^--X^+}\frac{1-\frac{1}{y^+X^-}}{1-\frac{1}{y^-X^+}}\frac{y^--X^-}{y^+-X^+}\frac{1-\frac{1}{y^-X^-}}{1-\frac{1}{y^+X^+}},
\eeq
$\sigma_{\rm BES}$ being the BES \cite{BES} dressing factor, and
\be
s_1(y^{\pm},X^{\pm})=1,\ s_2(y^{\pm},X^{\pm})=\frac{y^+-X^+}{y^+-X^-}\frac{1-\frac{1}{y^{-}X^{+}}}{1-\frac{1}{y^{-}X^{-}}},\ s_{3,4}(y^{\pm},X^{\pm})=\frac{y^+-X^+}{y^+-X^-}\sqrt{\frac{X^{-}}{X^{+}}}.
\label{elements}
\ee
Here we are using the usual kinematic variables for the virtual particle, solutions of the conditions
\be
\frac{y^-}{y^+}=e^{iq_{\star}};\quad y^{+}+\frac{1}{y^{+}}-y^{-}-\frac{1}{y^{-}}=\frac{i}{g},
\ee
and for the dyonic magnon:
\be
\frac{X^+}{X^-}=e^{ip};\quad X^{+}+\frac{1}{X^{+}}-X^{-}-\frac{1}{X^{-}}=\frac{iQ}{g}.
\ee

\subsection{Twisted algebraic curve and quantum finite-size correction from the $F$-term }
\newcommand{\adsiv}{${\rm AdS}_4/{\rm CFT}_3$}

\newcommand{\adsv}{${\rm AdS}_5/{\rm CFT}_4$}

\newcommand{\ba}{\begin{array}}

\newcommand{\ea}{\end{array}}
The (dyonic) giant magnon solution on the deformed $S^{5}_{\beta}$ can be described by the following set of twisted quasi-momenta
\begin{eqnarray}
{p}_{\hat1}(x)&=&  \frac{\alpha x}{x^{2}-1} + \phi_{\hat1};\ {p}_{\hat2}(x)=  \frac{\alpha x}{x^{2}-1} + \phi_{\hat2};\ {p}_{\hat3}(x)=  \frac{-\alpha x}{x^{2}-1} + \phi_{\hat3};\ {p}_{\hat4}(x)=  \frac{-\alpha x}{x^{2}-1} + \phi_{\hat4};\nonumber\\
{p}_{\tilde1}(x)&=&\frac{\alpha x}{x^{2}-1} +i \log\left(\frac{1/x- X^{+}}{1/x- X^{-}}\right)+ \phi_{\tilde1};\ {p}_{\tilde2}(x)=\frac{\alpha x}{x^{2}-1} -i \log\left(\frac{x-X^{+}}{x-X^{-}}\right)+ \phi_{\tilde2}\nonumber\\
{p}_{\tilde3}(x)&=&\frac{-\alpha x}{x^{2}-1} +i \log\left(\frac{x-X^{+}}{x-X^{-}}\right)+ \phi_{\tilde3};\ {p}_{\tilde4}(x)=\frac{-\alpha x}{x^{2}-1} -i \log\left(\frac{1/x- X^{+}}{1/x- X^{-}}\right)+ \phi_{\tilde4}\,,~~~~~~~~
\label{quasimomenta}
\end{eqnarray}
where $\alpha=\Delta/g$, $\Delta=J-Q+\frac{g}{i}(X^{+}-X^{-})$ and, since the deformation does not affect $AdS_{5}$, $\phi_{\hat1},...,\phi_{\hat4}=0$. The twists $\phi_{\tilde1},...,\phi_{\tilde4}$ can be fixed by observing that, in the language of \cite{GV1}, the twists $(\phi_{\tilde1},\phi_{\hat1},\phi_{\hat2},\phi_{\tilde2},\phi_{\tilde3},\phi_{\hat3},\phi_{\hat4},\phi_{\tilde4})$ correspond to $(\phi_{1},\phi_{2},\phi_{3},\phi_{4},\phi_{5},\phi_{6},\phi_{7},\phi_{8})$ \cite{GSV}, and then by comparing the twisted BAEs of \cite{GV1} to the Beisert-Roiban BAEs \cite{BeiRoi,ABBN2} with $\gamma_{1}=\gamma_{2}=\gamma_{3}=2\pi\beta$ , $L=J+Q$.
For giant magnon states, we set all the numbers of Bethe roots in the ``$SU(2)$'' grading to zero except the $SU(2)$ Bethe roots with $K_{4}\equiv Q$ and used the condition $\prod_{j=1}^{Q}\frac{x_j^+}{x_j^-}=e^{ip}$. 
Then the resulting twists are
\begin{eqnarray}
&\phi_{\tilde1}=p/2 + \pi\beta Q\,;\quad
\phi_{\tilde2}=-p/2 - \pi\beta Q\,;&\nonumber\\
&\phi_{\tilde3}=p/2 + \pi\beta (2L-3Q)\,;\quad
\phi_{\tilde4}=-p/2 - \pi\beta (2L-3Q)\ .&
\label{twists}
\end{eqnarray}
Another possible way is to use the twisted boundary conditions for the worldsheet excitations set by \cite{Frolov,Arutyunov:2010gu} \begin{equation}
Z \leftrightarrow e^{i2\pi\beta Q}\ ;\quad Y_{1\dot1}\leftrightarrow e^{i2\pi\beta J}\ ;\quad Y_{2\dot1}\leftrightarrow e^{i2\pi\beta (J-Q)}
\label{scalars}
\end{equation}
for the scalars, and
\begin{equation}
\theta_{1{\dot\alpha}}\leftrightarrow e^{i\pi\beta Q}\,;\quad \theta_{2{\dot\alpha}}\leftrightarrow e^{-i\pi\beta Q}\ ;\quad\eta_{\dot1\alpha}\leftrightarrow e^{i\pi\beta (2J-Q)}\,;\quad \eta_{\dot2\alpha}\leftrightarrow e^{-i\pi\beta (2J-Q)}
\label{fermions}
\end{equation}
for the fermions with $\alpha=3,4$.
Then one can obtain the twists (\ref{twists}), up to the terms depending on the momentum $p$, by
mapping the worldsheet excitations to the various physical polarizations of the algebraic curve fluctuations \cite{CDLM}:
\begin{eqnarray}
\label{polarizations}
&\hspace{-0.5cm}(ij)_{AdS_{5}}=(\hat1\hat3), (\hat1\hat4), (\hat2\hat3), (\hat2\hat4)\leftrightarrow (Z_{3\dot3},Z_{3\dot4},Z_{4\dot3},Z_{4\dot4});&\nonumber\\
&\hspace{-0.5cm}(ij)_{S^{5}}=(\tilde1\tilde3), (\tilde1\tilde4), (\tilde2\tilde3), (\tilde2\tilde4)\leftrightarrow (Y_{2\dot1},Y_{2\dot2},Y_{1\dot1},Y_{1\dot2});&\\
&\hspace{-0.5cm}(ij)_{Fermions}=(\hat1\tilde3), (\hat1\tilde4), (\hat2\tilde3), (\hat2\tilde4),(\tilde1\hat3), (\tilde1\hat4), (\tilde2\hat3), (\tilde2\hat4)\leftrightarrow (\eta_{\dot13},\eta_{\dot23},\eta_{\dot14},\eta_{\dot24},\theta_{2\dot3},\theta_{2\dot4},\theta_{1\dot3},\theta_{1\dot4}).&\nonumber
\end{eqnarray}
If we use $\tilde{\tilde\phi}_1(2\pi)-\tilde{\tilde\phi}_1(0)=p=p_{ws}+2\pi\beta Q$ and $\tilde{\tilde\phi}_2(2\pi)-\tilde{\tilde\phi}_2(0)=2\pi(n_2-\beta J)$ in the notations of \cite{FrRoTs}, our twists (\ref{twists}) also match the quasi-momentum asymptotic behaviors for the $SU(2)_{\beta}$ 
sector derived there 
\footnote{Actually it is not clear how
to extend the analysis of \cite{FrRoTs} 
to unphysical configurations, such as a single (dyonic) giant magnon, and to all the finite-gap solutions of the $\beta$-deformed theory.
We thank S.Frolov for making this point.}
\begin{equation*}
P(x)\mathop{\longrightarrow}_{x\rightarrow\infty}\frac{p_{ws}}{2}+\pi\beta (J+Q) -\frac{2\pi(J-Q)}{\sqrt{\lambda}x}+\ldots\ ;\quad P(x)\mathop{\longrightarrow}_{x\rightarrow 0}-\frac{p_{ws}}{2}+\pi\beta (J-Q) +\frac{2\pi(J+Q)}{\sqrt{\lambda}}x+\ldots.
\end{equation*}
where $P(x)=\frac{1}{2}(p_{\tilde3}(x)-p_{\tilde2}(x))=\frac{1}{2}(p_{\tilde1}\left(1/x\right)-p_{\tilde4}\left(1/x\right))$ \footnote{The twisted quasi-momenta (\ref{quasimomenta}) with the twists (\ref{twists}) satisfy the inversion symmetry $p_{\tilde1,\tilde2,\tilde3,\tilde4}(x)=-p_{\tilde2,\tilde1,\tilde4,\tilde3,}(1/x),\ p_{\hat1,\hat2,\hat3,\hat4}(x)=-p_{\hat2,\hat1,\hat4,\hat3,}(1/x)$.}.
 
While the twisted quasi-momenta are shifted by constants,
the fluctuation frequencies $\Omega_{ij}\left(x\right)$
of the deformed theory are the same as those of the undeformed theory and
polarization independent, i.e. same for all the $(i,j)$ \cite{GSV}:
\begin{equation}
\Omega_{ij}\left(x\right)=\frac{2}{x^{2}-1} \left(1-x \frac{X^{+}+X^{-}}{X^{+}X^{-}+1}\right).
\end{equation}
The one-loop quantum effects are the summation over all fluctuation frequencies,
\begin{equation}
\delta\Delta_{\rm one-loop}= \frac{1}{2}\sum_{ij}\sum_{n}\left(-1\right)^{F_{ij}}\Omega_{ij}^{n}
=\int \frac{dx}{2\pi i} \partial_{x}\Omega(x) \sum_{ij} \left(-1\right)^{F_{ij}}e^{-i\left(p_{i}-p_{j}\right)},\nonumber
\end{equation}
where the sum runs over all the physical polarizations (\ref{polarizations}).
The only change from the computations for the undeformed theory is the summand in the integral above, that is
\begin{eqnarray}
\sum_{ij}(-1)^{F_{ij}}e^{-i(p_{i}-p_{j})}&=&e^{-i\frac{2\alpha x}{x^2-1}}\left(e^{i\pi\beta (2J-Q)}\frac{x-X^{-}}{x-X^{+}}\sqrt{\frac{X^+}{X^-}}+e^{-i\pi\beta (2J-Q)}\frac{x X^{+}-1}{xX^{-}-1}\sqrt{\frac{X^-}{X^+}}-2\right)\nonumber\\
&\times&\left(e^{i\pi\beta Q}\frac{x-X^{-}}{x-X^{+}}\sqrt{\frac{X^+}{X^-}}+e^{-i\pi\beta Q}\frac{x X^{+}-1}{xX^{-}-1}\sqrt{\frac{X^-}{X^+}}-2\right) . \nonumber
\end{eqnarray}
For the non-dyonic giant magnon, one should take a limit $Q\to 1$ and then
$\beta Q\to0, X^{\pm}\to e^{\pm ip/2}$.

It can be shown explicitly that this result matches exactly the S-matrix supertrace given by Eqs. (\ref{summand}) and (\ref{elements}), once it is multiplied by the exponential factor $e^{-iq_{\star}J}\simeq e^{-i\frac{2Jx}{g(x^{2}-1)}}$, in the strong coupling approximation $y^{\pm}\simeq x$.
On the other hand, the matching of the kinematic part
\be
-\int_{\mathbb{R}}\frac{dq}{2\pi}\left(1-\frac{\epsilon_{Q}'(p)}{\epsilon_{1}'(q_{\star})}\right)...=\int_{U^{+}}\frac{dx}{2\pi i}\partial_{x}\Omega(x)...
\ee
is inherited without changes from the undeformed case \cite{GSV}.
This completes the matching and then confirms the validity of the quantum corrections calculated by using our $F$-term formula (\ref{fterm}) and the twisted quasimomenta (\ref{quasimomenta}).

\subsection{The $\mu$-term calculation}

In order to calculate explicitly the $\mu$-term from Eq. (\ref{muterm}), we shall follow basically the calculations of \cite{HatSuzi}. We just recall here that we need to compute the residues of the $S$-matrix (\ref{summand})-(\ref{elements}) in both its $s$-channel pole at $y^-=X^+$ and $t$-channel pole at $y^{+}=X^{+}$.
Then, since $s_{2}$, $s_{3}$ and $s_{4}$ are negligible in the classical limit $g>>1$,
we need to consider only the $s_1$ factors, multiplied by the respective twists $e^{i2\pi\beta J-Q}$ and $e^{i\pi\beta Q}$, which will give a final overall factor $e^{2i\pi\beta J}$ in front of the result of \cite{HatSuzi}.

Indeed, we have that, at both poles $y^-=X^+$ and $y^{+}=X^{+}$, the virtual particle momentum $q_{\star}$ and the exponential factor become
\be
{\tilde q}^*=-\frac{i}{g\sin\left(\frac{p-i\theta}{2}\right)}\quad \to\quad
e^{-i{\tilde q}^* J}\approx\exp\left[
-\frac{J}{g\sin\left(\frac{p-i\theta}{2}\right)}\right],
\ee
where we introduced $\theta$ defined by 
\be
\sinh\frac{\theta}{2}&\equiv&\frac{Q}{2g\sin\frac{p}{2}}.
\ee
From Eq.(\ref{disper}) one obtains
\be
1-\frac{\epsilon'_{Q}(p)}{\epsilon'_{1}({\tilde q}^*)}\approx
\frac{\sin\frac{p}{2}\sin\frac{p-i\theta}{2}}{\cosh\frac{\theta}{2}},
\ee
while the explicit evaluation of the residues at the leading order gives
\be
\frac{1}{(y^{\pm})'}\mathop{\rm Res}_{y^{\pm}=X^{+}}S_{0}^{2}=\pm\frac{4ig \sin^2\frac{p}{2}}{\sin\frac{p-i\theta}{2}}
e^{2\pi i\beta J}
\exp\left[-\frac{\epsilon_{Q}(p)}{g\sin\frac{p-i\theta}{2}}\right].
\ee
Combining all these contributions together, taking the difference of the contribution from the residue in $y^{-}=X^{+}$ and $y^{+}=X^{+}$ \cite{HatSuzi} and the real part of the final result, we get
\be
\delta E_{(1\dot1)_{Q}}^{\mu}&=&-\frac{8g\sin^3\frac{p}{2}}{\cosh\frac{\theta}{2}}
{\rm Re}\left\{e^{2\pi i\beta J}\
\exp\left[-\frac{J+\epsilon_{Q}(p)}{g\sin\frac{p-i\theta}{2}}\right]\right\}\non\\
&=&-\frac{16g^2\sin^4\frac{p}{2}}{\epsilon_Q(p)}\cos{\Phi}
\exp\left[-\frac{2\sin^2\frac{p}{2}\left[J+\epsilon_Q(p)\right]
\epsilon_Q(p)}{Q^2+4g^2\sin^4\frac{p}{2}}\right],
\ee
that agrees with Eq. (\ref{AhnBoz}), with $\Phi$ being exactly the same as Eq.(\ref{AhnBozi}).
In particular, in the non-dyonic limit $\theta\to 0$, the result reduces to
\be
\delta E_{(1\dot1)_{Q=1}}^{\mu}=-\frac{8g}{e^2} \sin^3\frac{p}{2}\cos(2\pi\beta J)
\exp\left[-\frac{J}{2g\sin\left(\frac{p}{2}\right)}\right],
\ee
that matches exactly Eq.(\ref{AhnBozii}).

\setcounter{equation}{0}
\section{Concluding Remarks}
In this note we have proposed L\"uscher formulas for $\mu$-term and $F$-term
corrections of a dyonic magnon state for the $\beta$-deformed $AdS_{5}/CFT_{4}$ theory. 

It turns out that the resulting finite-size corrections depend on the parameter $\beta$ only through an overall factor $\cos(2\pi\beta J)$, which has been observed for the first time in  \cite{AhnBoz} and  \cite{AhnBoz2}. The expression of the phase $\Phi$ is then in contrast to that derived in \cite{BykFro}, and has been confirmed in this letter both in the dyonic and non-dyonic cases, by classical and first quantum finite-size corrections calculated on the basis of the S-matrix proposed in \cite{ABBN2}, but we checked that the same results can be derived by using the Y-system's asymptotic solutions of \cite{GroLev} or the twisted transfer matrices derived by \cite{Arutyunov:2010gu}.
Then essentially we solved the long standing issue of matching string results for the finite-size effects of giant magnons on the $\beta$-deformed $S^{5}_{\beta}$ and L\"uscher corrections \cite{Zoubos,Janik}, that are derived by using the information of a twisted $S$-matrix with twisted BCs.

Now, it would be interesting to extend our analysis of the strong coupling finite-size corrections to all the orders in the volume $L$, along the lines of \cite{G}. This would entail the formulation and the solution of a set of twisted TBA/Y-system equations for $SU(2)$ excited states.
Also the analysis of the three-parameters deformation would be an interesting generalization of our results.

\section*{Acknowledgements}

We would like to thank In\^es Aniceto and Sergey Frolov for useful discussions and APCTP for the warm hospitality during the completion of this work. DB is also grateful to IEU for travel financial support and generous hospitality.\\
This work was supported in part by the WCU Grant No. R32-2008-000-101300 (CA),
the FCT fellowship SFRH/BPD/69813/2010 (DB), and by the NRF Grant No. 2005-0049409 (MK).

\end{document}